\newcommand{\al}{\alpha} 
\newcommand{\Om}{\Omega}\newcommand{\om}{\omega}
\newcommand{\ep}{\epsilon}
\newcommand{\eps}{\varepsilon} \newcommand{\NNn}{{\Bbb N}}
\newcommand{\de}{\delta}

\newcommand{\R}{{\Bbb R}} 
\newcommand{\pa}{\partial}
 \newcommand{\G}{{\cal G}}
\newcommand{\DD}{{\cal D}}\newcommand{\E}{{\cal E}}

\newcommand{\fr}{\frac{1}}
\newcommand{\nn}{\nonumber}
\newcommand{\ba}{\begin{array}}\newcommand{\ea}{\end{array}}
\newcommand{\beq}{ \begin{equation} }\newcommand{\eeq}{ \end{equation} }
\newcommand{\bea}{\begin{eqnarray}}\newcommand{\eea}{\end{eqnarray}}
\newcommand{\beas}{\begin{eqnarray*}}\newcommand{\eeas}{\end{eqnarray*}}
\newcommand{\beqn}{ \begin{equation*} }

\newtheorem{theorem}{Theorem}

\newtheorem{prf}{{\it Proof}}

\documentstyle[amsfonts]{article}
\title{On the geometry of impulsive gravitational waves}
\author{R. Steinbauer
	\footnote{Electronic mail: roland.steinbauer@univie.ac.at}\\
	\\{\normalsize Department of Mathematics, University of Vienna,  
	Strudlhofg.~4}\\{\normalsize A-1090 Wien, Austria}\\{\normalsize and}\\
	 {\normalsize Institute for Theoretical Physics, University of Vienna,
        Boltzmanng.~5}\\{\normalsize  A-1090 Wien, Austria}\\}
\date{September 14, 1998}
\begin{document}
\maketitle
\thispagestyle{empty}           

\begin{abstract}
We describe impulsive gravitational pp-waves entirely in the distributional 
picture. Applying Colombeau's nonlinear framework of generalized functions
we handle the formally ill-defined products of distributions which enter the 
geodesic as well as the geodesic deviation equation. Using a universal 
regularization procedure we explicitly derive regularization independent 
distributional limits. In the special case of impulsive plane waves we compare 
our results with the particle motion derived from the continuous form of the 
metric. 

{\em PACS-numbers: }04.20.Cv, 04.20.-q, 02.20.Hq, 04.30.-w 
\newline
\end{abstract}

\rightline{UWThPh -- 1998 -- 30}
\newpage        
\section{Introduction}
Plane fronted gravitational waves with parallel rays (pp-waves) are
spacetimes admitting a covariantly constant null vector field, 
which can be used to write the metric tensor in the form~\cite{ehlers+kundt}
\beq\label{distrm} ds^2\,=\,H(u,x,y)du^2-du\,dv+dx^2+dy^2,\eeq
where $u,v$ and $x,y$ is a pair of null and transverse (Cartesian) 
coordinates respectively.  
In this work we shall deal especially with {\em impulsive} pp-waves which can
be described by a profile function $H$ proportional to a 
$\de$-distribution~\cite{penrose}, i.e. $H(u,x,y)=f(x,y)\de(u)$,
where $f$ is assumed to be smooth. Hence the spacetime is flat everywhere, 
exept for the null hypersurface $u=0$, where it has a $\de$-shaped ``shock''. 
Such geometries arise physically as ultrarelativistic limits of boosted black 
hole spacetimes of the Kerr-Newman family~\cite{as,bn} and multipole solutions
of the Weyl family~\cite{pg}. Moreover they play an important 
role in particle scattering at the Planck scale~\cite{ls}.

There are also intrinsic descriptions of impulsive pp-waves, most prominently, 
Penrose's ``scissors and paste approach''~\cite{penrose}, which 
consists in glueing together two copies of Minkowski space along the null
hypersurface $u=0$, identifying points according to $(0,v,x,y)=(0,v+(1/2)f(x,y)
,x,y)$.
Penrose also introduced a different coordinate system in which the components of
the metric tensor are actually continuous and, in 
the special case of a {\em plane} wave, i.e. $f(x,y)=x^2-y^2$ take the form
\beq \label{contm}
	ds^2\,=\,-du\,dV+(1-u_+)^2dX^2+(1+u_+)^2dY^2,
\eeq
where $u_+=\theta(u)u$ denotes the ``kink-function'' and $\theta$ is the step
function. Clearly, the transformation relating (\ref{distrm}) with
(\ref{contm}) has to be discontinuous, i.e. is given by~\cite{penrose}
(for the case of a general pp-wave see~\cite{aib})
\bea\label{trsf}
 	x&=&(1+u_+)X 	\nn\\
	y&=&(1-u_+)Y	\\
	v&=&V+X^2(u+1)\theta(u)+Y^2(u-1)\theta(u)\nn
\eea 
Hence --strictly speaking-- the (topological) structure of the manifold is 
changed.

In this work we begin our investigations using the original distributional 
form of the metric, motivated by the fact that physically, i.e. in the 
ultrarelativistic limit, the spacetime arises that way 
(cf. also the approaches of~\cite{fvp,ba}).
Solving the geodesic as well as the geodesic deviation equation, we describe 
the geometry of impulsive pp-waves entirely in the distributional picture. 
As discussed in detail in~\cite{geo} these equations, due to their nonlinearity 
and the presence of the Dirac $\de$-function in the spacetime metric, 
involve formally ill-defined products of distributions. To overcome this problem
we use the setting of Colombeau algebras of generalized 
functions~\cite{co1,co2,moBOOK}, which we briefly describe in Sec.~\ref{ma}.

Finally, for the special case of plane impulsive waves, we compare our results
with calculations using the continuous coordinate system 
(see also~\cite{pv}) to find that the particle motion in both spacetimes 
agree. Hence we argue that from a physical point of view the spacetime 
manifolds are identical. 
\section{Colombeau Algebras}\label{ma}
In the vector space of distributions $\DD'$ no meaningful product can be 
defined. Moreover, as {\em L. Schwartz}~\cite{s} showed in 1954, there does not 
even exist an associative and commutative differential algebra containing 
the space of {\em $C^k$-functions} as a subalgebra that allows a linear 
embedding of distributions. 
{\em J.F. Colombeau}~\cite{co1,co2,moBOOK} 
introduced differential algebras $\G$
containing the space of distributions as a subspace, and the space of {\em
smooth }functions as a faithful subalgebra. In the light of Schwartz's so called
``impossibility result'' this framework has to be regarded as the best one can
hope for.

To  begin  with,  we give a short description of the algebra we
are going to use in the sequel which provides a natural
framework for studying nonlinear operations on singular data, hence singular 
(linear  and  nonlinear)  partial differential  equations.
The main idea is to regularize singular objects by sequences of smooth
functions, i.e. in the space $\E(\Omega):=C^\infty(\Omega)^{(0,1)}$, where 
$\Omega$ denotes an arbitrary open set of $R^n$. To achieve the consistency 
properties discussed above we introduce the spaces
\bea
{\cal E_M}(\Omega)
 &=& \{\,(u_\epsilon)_{\epsilon}\in(C^{\infty}(\Omega))^{(0,1)}:
 \,\forall K\subset\subset\Omega \, \forall \alpha\in \NNn_0^n\,\exists N>0:\nn
\\&&
\hskip1cm\sup\limits_{x\in K}\mid\partial^\alpha u_\epsilon(x)\mid=O
 (\epsilon^{-N})\quad(\epsilon\to 0)\,\}\,\,,\nopagebreak[0]
\\{\cal N}(\Omega)
 &=&\{\,(u_\epsilon)_{\epsilon}\in (C^\infty(\Omega))^{(0,1)}:
 \,\,\forall K\subset\subset\Omega \,\forall \alpha\in \NNn_0^n,\forall M>0:\nn
\nopagebreak[0]\\&&\hskip1cm
\sup\limits_{x\in K}\mid\partial^\alpha u_\epsilon(x)\mid
  =O(\epsilon^{M})\quad(\epsilon\to 0)\,\}\,.
\eea 
${\cal E_M}(\Om)$ is a differential algebra with pointwise operations and
${\cal N}(\Om)$ is an ideal in it. We define the {\em algebra of generalized
functions, }or {\em Colombeau algebra,} by the quotient
\beq \G(\Om)\,:=\,{\cal E_M}(\Om)\,/\,{\cal N}(\Om)\eeq 
and denote its elements by 
$ u\,=\,(u_\ep)_\ep\,+\,{\cal N}(\Om)$ or-more carelessly-by $u_\ep$.

Distributions with compact support are now embedded into
$\G(\R^n)$ by convolution with a {\em mollifier $\rho_\ep$, } defined as
follows; let $\rho\in {\cal S}(\R^n)$ (Schwartz's space) with the properties
$\int\rho(x)\,dx=1$ and $\int x^\alpha\rho(x)\,dx=0 \quad\forall \alpha\in
\NNn^n,\,\mid\!\al\!\mid\geq1$, then we set
$\rho_\ep(x):=(1/\ep^n)\rho(x/\ep)$. So we have the map
$\iota_0(\om)=(\om*\rho_\ep)_\ep+{\cal N}(\R^n)$.
This embedding can be
``lifted'' to an embedding $\iota: $ $\DD'(\Om)\hookrightarrow\G(\Om)$ by means
of sheaf theory while smooth functions are embedded as constant sequences,
i.e. $\sigma(f)=(f)_\ep$.

Next we briefly recall the concept of association in the Colombeau algebra,
which is essential in applications. Two generalized functions $u,v$ 
are called {\em associated} with each other ($u\approx v$),
provided their difference converges to $0$ in $\DD'$. 
In particular, if $u\in {\cal  D}'$ then it is 
called the macroscopic aspect (or  {\em distributional  shadow})  of  $v$.  
Equality  in $\DD'$ is reflected  as  equality  in  the sense of association 
in ${\cal G}$, while equality in ${\cal G}$ is a stricter concept (for
example, all  powers  of the Heaviside function are distinct in
the  Colombeau  algebra  although they are associated with each
other; moreover not every Colombeau function ``casts a distributional shaddow,''
for example $\iota(\de)^2$).

The construction introduced above is known as the ``special'' or ``simplified'' 
variant of Colombeau's algebra. A more refined construction, i.e. the ``full''
algebra, avoids the dependence on $\rho$ of the embedding of $\DD'$. At the 
expense of some technicalilties one can also construct a diffeomorphism 
invariant embedding. However, for the present application it will be sufficient to work 
in the ``special'' algebra.
\vskip12pt

In the mathematical literatur Colombeau algebras have been extensively used to
study nonlinear PDEs with singular data or coefficients (see eg.~\cite{moBOOK}
and the references therein). Recently there have also been a number of
applications to general relativity, in particular the calculation of the
curvature of cosmic strings~\cite{string1,string2,string3} and the
ultrarelativistic limits of Kerr-Newman black holes (see \cite{ul1,ul2} and
references therein). For a current overview of the topic see the review
article of J. Vickers~\cite{vickers}. 
\section{Solving the Geodesic and Geodesic Deviation Equation in ${\bf{\cal
G}}$} \label{es}
We start with a pp-metric of the form \beq
\label{pp}ds^2\,=\,f(x^i)\,\de(u)\,du^2-du\,dv+(dx^i)^2\,\,, \eeq where $x^i$
($i=1,2$) denote the transverse coordinates.  It is straightforward to derive
the geodesic equation which (using $u$ as an affine parameter, thereby
excluding only trivial geodesics parallel to  the shock) takes the form
\bea\label{geo}      \ddot v(u)&=&f(x^j(u))\,\dot\de(u)              
\,+\,2\,\pa_i\,f(x^j(u))\,\,\dot x^i\,(u)\,\de(u)\,\,,              \nn \\    
 \ddot x^i\,(u)&=&\frac{1}{2}\,\pa_i\,f(x^j(u))\,\,\de(u)\,\,, \eea where
$\dot{}$ denotes the derivative with respect to $u$. 

Note that the first line involves a product of the $\de$-function with $\dot
x^i$ which we cannot even expect to be continuous (and in fact turns out to be
proportional to the Heaviside function, cf.~\cite{geo}). We proceed by
transferring the geodesic equations into Colombeau's framework.
The general
strategy for solving differential equations in $\G$ is to embed 
singularities (in our case: $\de$) into $\G$, which amounts to a
regularization, and  then  solve  the corresponding regularized
equations. In order to obtain general results we are therefore 
interested    in  imposing  as  few restrictions as possible on
the regularization of $\de$. The largest ``reasonable'' class of 
smooth\footnote{Note that, since ${\cal D}$ is dense in $L^1$, 
practically even discontinuous regularizations (eg. boxes) are included.}
regularizations  of $\de$ is given by nets 
$(\rho_\eps)_{\eps\in (0,1)}$ of smooth functions $\rho_\eps$ 
satisfying:
\bea
\mbox{(a)}
&\quad&   \mbox{\rm   supp}(\rho_\eps)  \to  \{0\}  \quad (\eps\to  0)\,\,,
\nn\\
\mbox{(b)}
&\quad& \int \rho_\eps (x) \,dx \to 1 \quad (\eps\to 0)\,\,\mbox{ and} \nn\\
\mbox{(c)}
&\quad& \int |\rho_\eps (x)| \,dx 
\mbox{ is bounded uniformly for small } \ep \nn
\eea
(cf. the definition of {\em strict delta  nets}  in \cite{moBOOK},  
ch.  2.7). Obviously  any  such net converges to $\delta$ in 
distributions as $\eps\to 0$.
To simplify notations it is often convenient to replace (a)
by
$$ \mbox{(a')}\quad
\mbox{\rm  supp}(\rho_\eps)  \subseteq  [-\eps,\eps] \;\;\;\;
\forall \eps\in (0,1).
$$
We shall call a net satisfying conditions (a'), (b) and (c) a {\em generalized 
delta function}.
Denoting the $\G$-functions corresponding to $x^i$ and $v$ by 
$x^i_\ep$ and $v_\ep$ we are now prepared to state the following
(cf.~\cite{geo2}) 
\begin{theorem} \label{geoth}
Let   $\rho_\ep\in  \G(\R)$  be  a  generalized  delta function. The regularized
geodesic equation
\bea\label{georeg}
     \ddot v_\ep(u)&=&f(x^j_\ep(u))\,\dot\rho_\ep(u)
              \,+\,2\,\pa_i\,f(x^j_\ep(u))\,\,\dot x^i_\ep(u)
              \,\rho_\ep(u)\,\,,
             \nn \\
     \ddot x^i_\ep(u)&=&\frac{1}{2}\,\pa_i\,f(x^j_\ep(u))\,\rho_\ep(u)
\eea
with initial conditions\footnote{Note that we have to impose initial conditions
``long before'' the shock. Geodesics starting at the shock cannot be treated
in a regularization independent manner.}
$v_\ep(-1)=v_0$, $\dot v_\ep(-1)=\dot v_0$, $x^i_\ep(-1)=x^i_0$, $\dot x^i_
\ep(-1)=\dot x^i_0$ has a unique locally bounded solution $(v_\ep,x^i_\ep)\in 
\G(\R)^3$. Moreover these solutions satisfy the following association relations
\bea
x^i_\ep &\approx & x_0^i + \dot x_0^i (1+u) + \frac{1}{2}\pa_i 
f(x_0^i + \dot x_0^i) u_+ \nn\\
v_\ep &\approx & v_0 + \dot v_0 (1+u) + f(x_0^i + \dot x_0^i) \theta(u)
+ \\
&\hphantom{\approx}& 
+\pa_if(x_0^i + \dot x_0^i) \left(\dot x_0^i + 
\frac{1}{4}\pa_if(x_0^i +  \dot x_0^i)\right) u_+ \,\,.\nn\label{geoass}
\eea
\end{theorem}

Hence from the distributional point of view the geodesics are given by refracted,
broken straight lines. More
precisely, the geodesics suffer a jump and a kink in the $v$- as well as a kink
in the $x^i$-direction when crossing the shock hypersurface. The scale of the
effect is entirely determined by the values of the profile function 
and its first derivatives at the shock hypersurface and (in the special case of
plane waves) reproduces exactly Penrose's junction conditions (\ref{trsf}). 
Note, however, that the above
results are regularization independent even within the maximal 
class of regularizations of the Dirac $\de$. 
\vskip12pt

Our  next  goal  is  an  analysis  of  the  geodesic deviation equation for
impulsive  pp-waves in the framework of algebras of generalized
functions.   As   in   \cite{geo} to keep formulas more transparent   
we  make  some  simplifying
assumptions  concerning  geometry (namely axisymmetry) and
initial  conditions. Writing $x=x^1$ and $y=x^2$ we  suppose  
that  $f$ depends
exclusively  on the two-radius $\sqrt{x^2+y^2}$ and work within
the hypersurface $y=0$ (corresponding to initial conditions 
$y_0=0=\dot  y_0$).  Furthermore we demand
$v_0=0=\dot x_0$. In this situation the Jacobi equation for the regularized 
deviation vector field $N^a_\ep(u)=(N^u_\ep(u),N^v_\ep(u),N^x_\ep(u),
N^y_\ep(u))$ takes the form
\bea\label{rj}
     \ddot N^u_\ep &=&0\qquad\ddot N^y_\ep \,=\,0\qquad
     \ddot N^x_\ep \,=\,[\dot {N^u_\ep} f'(x_\ep )+
                         \fr{2}\,N^x_\ep f''(x_\ep )]\rho_\ep 
                         +\fr{2}\,f'(x_\ep )N^u_\ep \dot\rho_\ep\nn\\
     \ddot N^v_\ep &=&2[N^x_\ep f'(x_\ep )\rho_\ep]\,\dot{\,}-
                        N^x_\ep f'(x_\ep )\dot\rho_\ep +
                        [N^u_\ep f(x_\ep )\rho_\ep]\,\ddot{\,}\\
                    &&\hskip2.3cm -N^u_\ep f''(x_\ep )\dot x_\ep^2 \rho_\ep -
                        N^u_\ep f'(x_\ep )\ddot x_\ep \rho_\ep\,\,, \nn
\eea
where $'$ denotes derivatives with respect to $r$ and we have suppressed the
parameter $u$. Furthermore $x_\ep$ is determined by  (\ref{georeg}) with
simplifiactions as discussed above.   Note that the last equation in (\ref{rj})
involves terms proprotional to $\theta^2\de$ and even $\de^2$. However,
existence  and  uniqueness of solutions to the corresponding  initial  value  
problem in the Colombeau algebra as well as (regularization independent!)
association relations (which we give explicitly for some special choices 
of initial conditions) are still guaranteed.
\begin{theorem} \label{jacth}
The Jacobi equation (\ref{rj}) with initial conditions $N^a_\ep(-1)=n^a$ and 
$\dot N^a_\ep(-1)=\dot n^a$ has a unique solution in $\G(\R)^4$.
If $N^a_\ep(-1)=0$ and $\dot N^a_\ep(-1)=(a,b,0,0)$ the unique solution
satisfies    the    following    association relations
\bea
N^x_\ep&\approx & \frac{1}{2} a f'(x_0)(u_++\theta(u)) \label{nxass}\nn\\
N^v_\ep &\approx & b(1+u) + a[f(x_0)\de(u) +\frac{1}{4}f'(x_0)^2
(\theta(u)+u_+)]\,. \label{nvass}
\eea
For initial conditions $N^a_\ep(-1)=(0,0,a,0)$, $\dot N^a_\ep(-1)=0$
we have
\bea\label{2nd}
N^x_\ep&\approx&a(1+u_+)\nn\\
N^v_\ep&\approx&af'(x_0)(\theta+\frac{1}{2}f''(x_0)u_+)\,.
\eea
\end{theorem}

Hence, viewed distributionally, in the first case the Jacobi field suffers a 
kink, a jump and a
$\de$-like pulse in the $v$-direction as well as a kink and jump in the 
$x$-direction overlapping the linear flat space behavior. These effects can be
understood heuristically from the corresponding behavior of the geodesics, 
given by equation~(\ref{geoass}). The constant factor $a$, which gives the
``scale'' of all the nonlinear effects, arises from the ``time advance'' 
of the ``nearby'' geodesics, represented by the initial velocity of the Jacobi
field in the $u$-direction. 
However the second case shows that this is not the only effect producing kinks
and jumps and we will refer to it in the discussion of the next Section.

\section{Discussion}\label{mf}

We have shown that the geometry of impulsive pp-waves can be described entirely 
within the distributional picture. Note that it was essential to use
regularization techniques, i.e. Colombeau's algebra, to handle the ill-defined
singular terms instead of introducing ``ad-hoc'' multiplication rules into
Schwartz linear theory. Even within the maximal class of regularizations of the
Dirac-$\de$ we were able to derive regularization independent distributional
geodesics and deviation fields. This (mathematical) feature also has a 
remarkable physical consequence. Interpreting the impulsive wave as a 
limiting case of sandwich waves of the form (cf.~\cite{rindler})
$H(u,x,y)=f(x,y)\rho_\ep(u)\to f(x,y)\de(u)$ we have shown that the impulsive
limit is totally independent of the special form of the original profile 
$\rho_\ep$ (see also the results in~\cite{pv1}).  

Finally (in the special case of plane waves) we discuss the relations 
of our approach to the one using the continuous form of the metric.
The metric (\ref{contm}) has the advantage that simple particle motion can bee
seen directly. Indeed free particles at fixed values of $X,Y$ and $Z=V-T$ 
after the shock start to move such that their relative $X$- and
$Y$-distance is given by the functions $1+u_+$ and $1-u_+$ respectively. This is
in total agreement with equation (\ref{2nd}).
(Note that it is the coordinate transformation which introduces the motion in 
$v$-direction.)
 
Moreover, we can solve the geodesic equations for the metric (\ref{contm}) either
by using the method of Sec.~\ref{es} or (since these equations only involve
Heaviside and kink functions) by solving them separatly for $u<0$ and 
$u>0$ and joining them in a $C^1$-manner. Either way leads to the
solutions ($u<1$, and using analogous initial values as before)
\bea X(u)&=&x_0+\dot x_0(2+u_-)-\frac{\dot x_0}{1+u_+}\nn\\
     Y(u)&=&y_0+\dot y_0u_-+\frac{\dot y_0}{1-u_+}\label{Geo}\\
     V(u)&=&v_0+\dot v_0(1+u)+\frac{\dot y_0^2u_+^2}{1-u}
     			    -\frac{\dot x_0^2u_+^2}{1+u}\nn\,\,,
\eea     
where $u_-:=\theta(-u)u$.

If we now formally transform equations (\ref{Geo}) according to (\ref{trsf})
we  again obtain the geodesics (\ref{geoass}).
Therefore we conclude that {\em physically} the two approaches 
to impulsive plane waves, hence the two differential structures of the
manifold,
are equivalent. However, the transformation once more involves products of
distributions ill-defined in the linear theory. Future work will be
concerned with a mathematical analysis and interpretation of this 
discontinuous change of coordinates. Note that the transformation 
(\ref{trsf}) is given precisely by the limit of the geodesic equations (\ref{geoass})
with vanishing initial velocities. Hence we can study the respective
transformation in the sandwich case and then approach the impulsive limit.

\section*{Acknowledgement}
The author wishes to thank D. Vulcanov for the kind invitation and hospitality
during the conference,
M. Kunziger for his collaboration and M. Oberguggenberger for so many helpful 
discussions. This work was supported by Austrian Academy of 
Science, Ph.D. programme, grant \#338 and by Research Grant P12023-MAT 
of the Austrian Science Foundation (FWF).
                 
\end{document}